\documentclass{PoS}

\title{Azimuthal correlations of forward di-hadrons in d+Au collisions suppressed by saturation}

\ShortTitle{Azimuthal de-correlations of forward di-hadrons}

\author{Javier L. Albacete\\
        Institut de Physique Th\'eorique, CEA/Saclay, 91191 Gif-sur-Yvette cedex, France\\
        E-mail: \email{javier.lopez-albacete@cea.fr}}

\author{\speaker{Cyrille Marquet}\\
        Physics Department, Theory Unit, CERN, 1211 Gen\`eve 23, Switzerland\\
        E-mail: \email{cyrille.marquet@cern.ch}}

\abstract{

RHIC experiments have recently measured the azimuthal correlation function of forward di-hadrons. The data show a disappearance of the away-side peak in central d+Au collisions, compared to p+p collisions, as was predicted by saturation physics. Indeed, we argue that this effect, absent at mid-rapidity, is a consequence of the small-$x$ evolution into the saturation regime of the Gold nucleus wave function. We show that the data are well described in the Color Glass Condensate framework.

}

\FullConference{35th International Conference of High Energy Physics - ICHEP2010,\\
		July 22-28, 2010\\
		Paris France}

\begin{document}

Hard processes in hadronic collisions, which resolve the partonic structure of hadrons, are well described by the leading-twist approximation of QCD. In this weak-coupling regime, partons in the hadronic wave function scatter independently, this is the essence of collinear factorization. However, since the parton densities grow with decreasing energy fraction $x,$ the hadronic wave function also features a low-$x$ part where the parton density has become large and partons scatter coherently, invalidating the leading-twist approximation. This weak-coupling regime, where non-linearities are important, is called saturation, and it can be probed at high-energies since increasing the energy of a collision allows to probe lower-energy partons.

In hadron-hadron collisions, by contrast with deep inelastic scattering, hard processes are singled out by requiring that one or more particles are produced with a large transverse momentum, much bigger than $\Lambda_{QCD}$. When in addition the particles are produced at forward rapidities, such processes are sensitive only to high-momentum partons inside one of the colliding (dilute) hadron, whose QCD dynamics is well understood, while mainly small-momentum (small-$x$) partons inside the other dense hadron contribute to the scattering. Replacing that hadron by a large nucleus further enhances the gluon density, and the possibility to reach the saturation regime.

In the case of single-inclusive hadron production, the suppression of particle production at forward rapidities in d+Au collisions compared to p+p collisions, experimentally observed at RHIC \cite{RdAu}, constitutes one of the most compelling indications for the presence of non-linear QCD evolution effects in presently available data. The Color Glass Condensate (CGC) provides a robust theoretical framework to describe the small-$x$ degrees of freedom of hadronic/nuclear wave functions. The good description of, among other observables, forward hadron production in d+Au collisions at RHIC \cite{Albacete:2010bs} indeed lends support to the idea that saturation effects may be a relevant dynamical ingredient at present energies.

However, alternative explanations of the suppressed forward hadron yield in d+Au collisions were proposed \cite{alternatives}, suggesting that one is not yet sensitive to the saturation of the nuclear gluon density at RHIC energies, but that the suppression is rather due to partonic energy loss through the nuclear matter, neglected in CGC calculations. In spite of the fact that saturation-based approaches were the only ones to correctly predict this phenomenon, the existence of alternative scenarios calls for the study of more complex observables. In the light of recent preliminary data in d+Au collisions at RHIC, showing the production of forward {\it mono-jets} \cite{rhicprel}, calculating double-inclusive forward hadron production in both frameworks could help pin down which is the correct picture. In this work, we show that the CGC calculation predicts correctly the azimuthal de-correlation of forward di-hadrons in d+Au collisions compared to p+p collisions \cite{Albacete:2010pg}, thus providing further support for the presence of saturation effects in present data.

\section{Formulation}

In the case of double-inclusive hadron production, denoting $p_{1\perp},$ $p_{2\perp}$ and $y_1,$ $y_2$ the transverse momenta and rapidities of the final state particles, the partons that can contribute to the cross section have a fraction of longitudinal momentum bounded from below, by $x_p$ (for partons from the deuteron wave function) and $x_A$ (for partons from the nucleus wave function), which are given by
\begin{equation}
x_p=x_1+x_2\ ,\quad
x_A=x_1\ e^{-2y_1}+x_2\ e^{-2y_2}\ ,
\quad\mbox{with}\quad
x_i=\frac{|p_{i\perp}|}{\sqrt{s}}\ e^{y_i}\ .
\label{kin}
\end{equation}
The kinematic range for forward particle detection at RHIC is such that, with $\sqrt{s}=200$ GeV, $x_p\!\sim\!0.4$ and $x_A\!\sim\!10^{-3}.$ Therefore the dominant partonic subprocess is initiated by valence quarks in the deuteron and, at lowest order in $\alpha_s,$ the $dAu\!\to\!h_1h_2X$ cross-section is obtained from the $qA\to qgX$ cross-section, the valence quark density in the deuteron $f_{q/d}$, and the appropriate hadron fragmentation functions $D_{h/q}$ and $D_{h/g}$:
\begin{eqnarray}
dN^{dAu\to h_1 h_2 X}=\int_{x_1}^1 dz_1 \int_{x_2}^1 dz_2 \int_{\frac{x_1}{z_1}+\frac{x_2}{z_2}}^1 dx\ \left[dN^{qA\to qgX}\left(xP,\frac{p_1}{z_1},\frac{p_2}{z_2}\right)D_{h_1/q}(z_1,\mu)D_{h_2/g}(z_2,\mu)+\right.\nonumber\\\left.
dN^{qA\to qgX}\left(xP,\frac{p_2}{z_2},\frac{p_1}{z_1}\right)D_{h_1/g}(z_1,\mu)D_{h_2/q}(z_2,\mu)\right]f_{q/d}(x,\mu)\ .
\label{collfact}
\end{eqnarray}
Here we will use the CTEQ6 NLO quark distributions and the KKP NLO fragmentation functions. The factorization and fragmentation scales are both chosen equal to the transverse momentum of the leading hadron, which we choose to denote hadron 1, $\mu=|p_{1\perp}|.$ Note that we have assumed that the two final-state hadrons come from partons which have fragmented independently, therefore formula (\ref{collfact}) cannot be used when $R^2=(y_2-y_1)^2+(\Delta\phi)^2$ is too small, where $\Delta\phi$ is the difference between the hadrons azimuthal angles. Computing the cross section for small $R$ requires the introduction of poorly-known di-hadron fragmentation functions, we shall not do it, because as we shall see the non-linear QCD effects we are interested in manifest themselves around $\Delta\Phi=\pi.$

As usual, due to parton fragmentation, the values of $x$'s probed are generically higher than $x_p$ and $x_A$ defined in (\ref{kin}). For the proton, one has $x_p<x<1$, and if $x_p$ would be smaller (this will be the case at the LHC), the gluon initiated processes $gA\to q\bar{q}X$ and $gA\to ggX$ should also be included in (\ref{collfact}), they have been computed recently \cite{Dominguez:2011wm} . For the nucleus, we shall see that the parton momentum fraction varies between $x_A$ and $e^{-2y_1}+e^{-2y_2}$. Therefore with large enough rapidities, only the small-$x$ part of the nuclear wave function is relevant when calculating the $qA\to qgX$ cross section, and that cross section cannot be factorized further: $dN^{qA\to qgX}\neq f_{g/A}\otimes dN^{qg\to qg}$. Indeed, when probing the saturation regime, $dN^{qA\to qgX}$ is expected to be a non-linear function of the nuclear gluon distribution, which is itself, through evolution, a non-linear function of the gluon distribution at higher $x$.

Using the CGC approach to describe the small-$x$ part of the nucleus wave function, the $qA\to qgX$ cross section was calculated in \cite{othercalc,Marquet:2007vb}. It was found that the nucleus cannot be described by only the single-gluon distribution, a direct consequence of the fact that small-$x$ gluons in the nuclear wave function behave coherently, and not individually. The $qA\!\to\!qgX$ cross section is instead expressed in terms of correlators of Wilson lines (which account for multiple scatterings), with up to a six-point correlator averaged over the CGC wave function, while the gluon distribution is the Fourier transform of only a two-point (dipole) correlator ${\cal N}$:
\begin{equation}
F(x,k_\perp)=\int\frac{d^2r}{(2\pi)^2}\ e^{-ik_\perp\cdot r}\ [1-{\cal N}(x,r)]\ ,
\label{ugd}
\end{equation}
where $r$ denotes the dipole transverse size. $F(x,k_\perp)$ is actually called the unintegrated gluon distribution, due to the fact that it is $k_\perp$ dependent, a feature known to be necessary to describe small-$x$ partons, even in the linear regime.

At the moment, it is not known how to practically evaluate the six-point function. In \cite{Marquet:2007vb}, an approximation was made which allows to express the higher-point correlators in terms of the two-point function (\ref{ugd}). This is done assuming a Gaussian distribution of the color sources, with a non-local variance. The resulting cross section for the inclusive production of the quark-gluon system in the scattering of a quark with momentum $xP^+$ off the nucleus $A$ reads \cite{Marquet:2007vb}:
\begin{equation}
\frac{dN^{qA\to qgX}}{d^3kd^3q}=
\frac{\alpha_S C_F}{4\pi^2}\ \delta(xP^+\!-\!k^+\!-\!q^+)\ F(\tilde{x}_A,\Delta)
\nonumber\\\sum_{\lambda\alpha\beta}
\left|I^{\lambda}_{\alpha\beta}(z,k_\perp\!-\!\Delta;{\tilde{x}_A})\!-\!
\psi^{\lambda}_{\alpha\beta}(z,k_\perp\!-\!z\Delta)\right|^2\ ,
\label{cs}
\end{equation}
where $q$ and $k$ are the momenta the quark and gluon respectively, and with $\Delta=k_\perp+q_\perp$ and $z=k^+/xP^+$. In this formula, $\tilde{x}_A$ denotes the longitudinal momentum fraction of the gluon in the nucleus, and $\tilde{x}_A=x_1\ e^{-2y_1}/z_1+x_2\ e^{-2y_2}/z_2>x_A$ when the cross section (\ref{cs}) is plugged into formula (\ref{collfact}).

The second line of formula (\ref{cs}) features the so-called $k_T$-factorization breaking term, with
\begin{equation}
I^{\lambda}_{\alpha\beta}(z,k_\perp;x)
=\int d^2q_\perp \psi^{\lambda}_{\alpha\beta}(z,q_\perp) F(x,k_\perp\!-\!q_\perp)\ ,
\label{split}\end{equation}
and where $\psi^{\lambda}_{\alpha\beta}$ is the well-known amplitude for $q\!\to\!qg$ splitting ($\lambda,$ $\alpha$ and $\beta$ are polarization and helicity indices). While no additional information than the two-point function is needed to compute (\ref{cs}), since higher-point correlators needed in principle have been expressed in terms of $F(x,k_\perp)$, the cross section is still a non-linear function of that gluon distribution, invalidating $k_T$-factorization. The rather simple form of the $k_T$-factorization breaking term is due to the use of a Gaussian CGC color source distribution, and to the large-$N_c$ limit. However, the validity of this approximation has been critically examined \cite{Dumitru:2010ak}, as it does not allow to correctly implement the non-linear QCD evolution, even in the large $N_c$ limit. Finally, let us comment on the factor $\delta(xP^+\!-\!k^+\!-\!q^+)$ in formula (\ref{cs}). This delta function is a manifestation of the fact that in a high-energy hadronic collision, the momentum transfer is mainly transverse, and it appears because the eikonal approximation was used to compute the $qA\to qgX$ cross section. This is valid in the high-energy limit, as for instance the energy loss of the incoming quark is neglected.

The CGC is endowed with a set of non-linear evolution equations which in the large-$N_c$ limit reduce to the Balitsky-Kovchegov (BK) equation \cite{loBK}. These equations can be interpreted as a renormalization group equation for the $x$ evolution of the unintegrated gluon distribution, and more generally of $n$-point correlators, in which both linear radiative processes and non-linear {\it recombination} effects are included. In this work, we compute the small-$x$ dynamics of the dipole correlator ${\cal N}$ by solving the running-coupling (rc) BK equation. The evolution kernel is evaluated according to the prescription of Balitsky. Explicit expressions, together with a detailed discussion on the numerical method used to solve the rcBK equation can be found in \cite{rcBK}, along with detailed discussions about other prescriptions proposed to define the running-coupling kernel.

The only piece of information left to fully complete all the ingredients in (\ref{collfact}) are the initial conditions for the rcBK evolution of ${\cal N}(x,r)$. This non-perturbative input has been constrained by single-inclusive forward hadron production data in \cite{Albacete:2010bs}. The two parameters are $x_0=0.02$, the value of $x_A$ below which one starts to trust, and therefore use, the CGC framework, and the value of the saturation scale at the starting point of the evolution $\bar{Q}_{s}^2(x_0)\equiv\bar{Q}^2_{s0}=0.4$ GeV$^2$. Then, this information was simply taken over in \cite{Albacete:2010pg}. In this respect, these forward di-hadron calculations are predictions, there are no free parameter to play with.

\section{Comparison with data}

We will now investigate the process $dAu\!\to\!h_1h_2X,$ with $\sqrt{s}\!=\!200\ \mbox{GeV}.$ In particular we shall study the $\Delta\phi$ dependence of the spectrum, where $\Delta\phi$ is the difference between the azimuthal angles of the measured forward particles $h_1$ and $h_2.$ To be more specific, we shall compute the coincidence probability to, given a trigger particle in a certain momentum range, produce an associated particle in another momentum range. It is given by 
\begin{equation}
CP(\Delta\phi)=\frac{N_{pair}(\Delta\phi)}{N_{trig}}\ ,\quad\mbox{with}\quad
N_{pair}(\Delta\phi)=\int\limits_{y_i,|p_{i\perp}|}
\frac{dN^{pA\to h_1 h_2 X}}{d^3p_1 d^3p_2}\ ,\quad
N_{trig}=\int\limits_{y,\ p_\perp}\frac{dN^{pA\to hX}}{d^3p}\ .
\end{equation}
In order to compare with the STAR measurement, the integration bounds for the rapidities are set to $2.4<y<4,$ which also ensures that only small-momentum partons are relevant in the nucleus wave function. In addition, for the trigger (leading) particle $|p_{1\perp}|>2$ GeV and for the associated (sub-leading) hadron $1\ \mbox{GeV}<|p_{2\perp}|<|p_{1\perp}|.$ The single-inclusive hadron production spectrum, used to normalize the coincidence probability, is calculated as explained in \cite{Albacete:2010bs}.

\begin{figure}
\includegraphics[width=7.5cm]{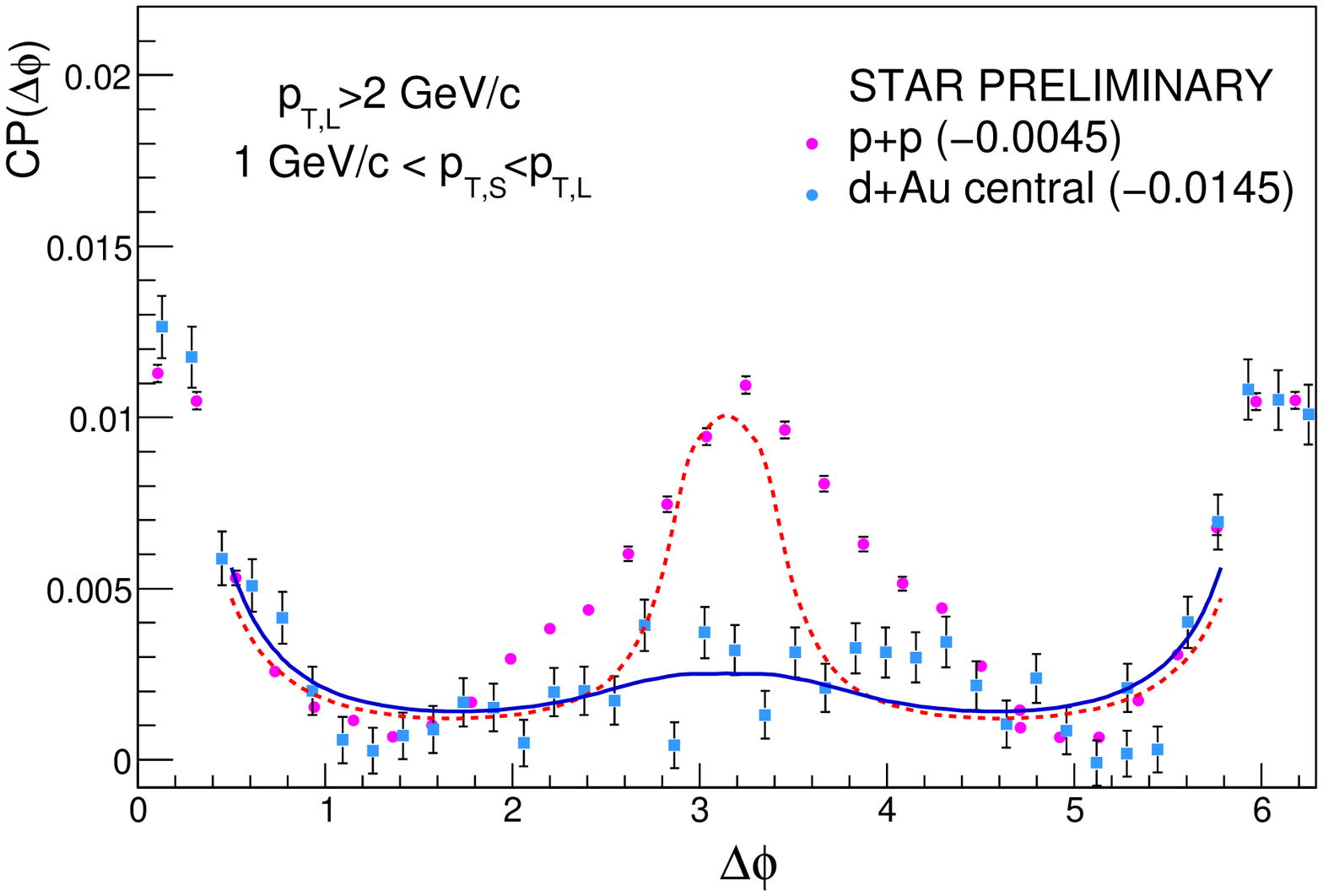}
\hfill
\includegraphics[width=7.5cm]{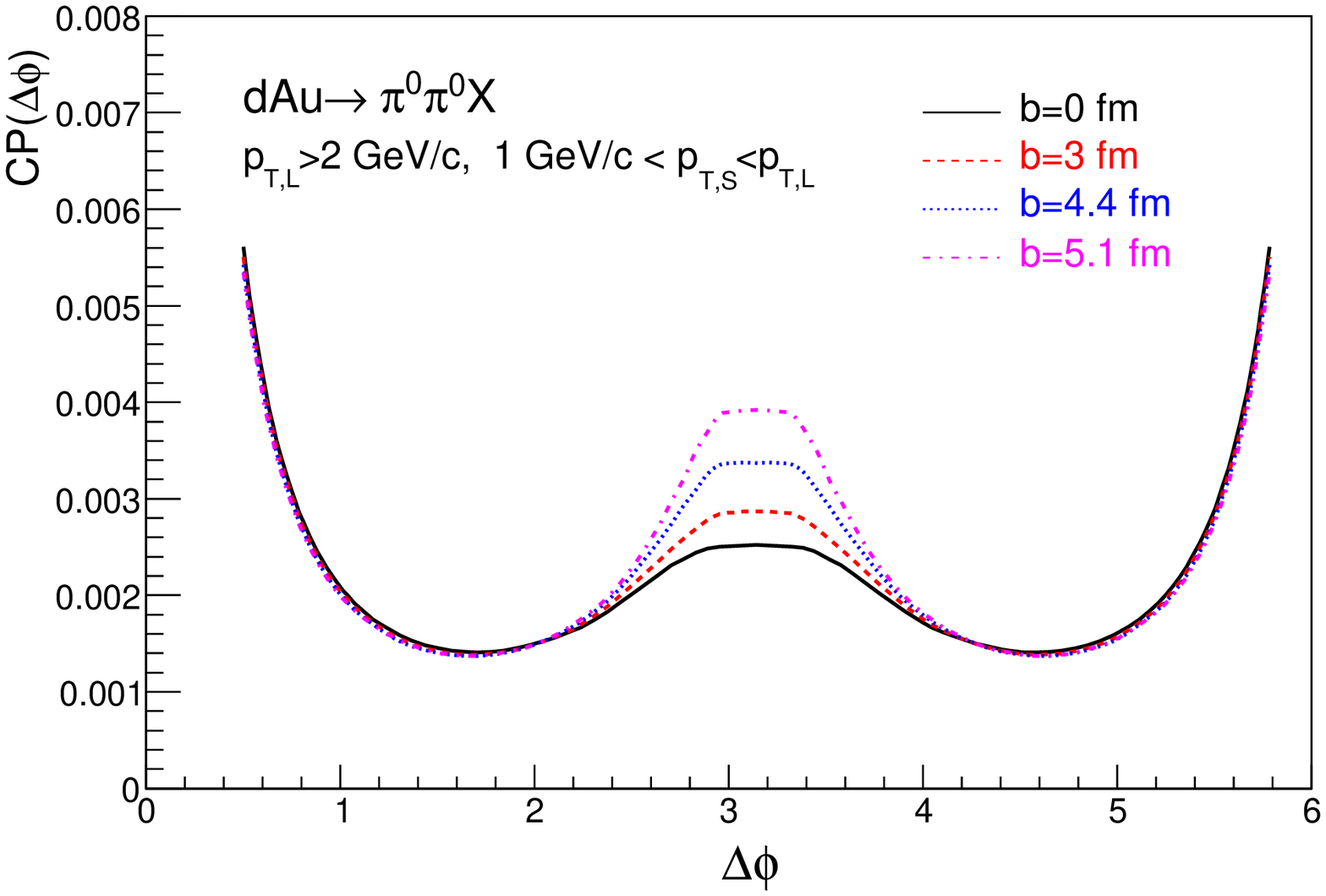}
\caption{The coincidence probability at a function of $\Delta\phi$. Left: CGC calculations \cite{Albacete:2010pg} for p+p and central d+Au collisions, the disappearance of the away-side peak is quantitatively consistent with the STAR data. Right: CGC predictions for different centralities of the d+Au collisions, the near-side peak is independent of the centrality, while the away-side peak reappears as collisions are more and more peripheral.}
\label{fig}
\end{figure}

To deal with the centrality dependence, we identify the centrality averaged initial saturation scale $\bar{Q}^2_{s0}$, extracted from minimum-bias single-inclusive hadron production data, with the value of $Q^2_{s0}$ at $b=5.47$ fm, and use the Woods-Saxon distribution $T_A(b)$ to calculate the saturation scale at other centralities: 
\begin{equation}
Q^2_{s0}(b)=\frac{\bar{Q}_{s0}^2\ T_A(b)}{T_A(5.47\ \mbox{fm})}\ ,\quad\bar{Q}^2_{s0}=0.4\ \mbox{GeV}^2\ .
\end{equation}
The result for central d+Au collisions is displayed in Fig.1, left plot, along with preliminary data from the STAR collaboration. As mentioned before, we do not calculated the complete near-side peak, as our formula does not apply around $\Delta\phi=0$. We see that the disappearance of the away-side peak in central d+Au collisions, compared to p+p collisions, is quantitatively consistent with the CGC calculations. The latter are only robust for central d+Au collisions, but the extrapolation to p+p collisions is displayed in order to show that it is qualitatively consistent with the presence of the away-side peak, and also with the fact that the near-side peak is identical in the two cases and is not sensitive to saturation physics. Since uncorrelated background has not been extracted from the data, the overall normalization of the data points has been adjusted by subtracting a constant shift.

In Fig.1, right plot, we show the centrality dependence of the coincidence probability. Although it is difficult to trust our formalism all the way to peripheral collisions, we predict that the near-side peak does not change with centrality, and that the away-side peak reappears for less central collisions. This is consistent with the fact that peripheral d+Au collisions are p+p collisions. The fact that the away-side peak disappears from peripheral to central collisions shows that indeed this effect is correlated with the nuclear density. Moreover di-hadron correlations at mid-rapidity, which are sensitive to larger values of $x_A$, feature an away-side peak whatever the centrality. The fact that for central collisions the away-side peak disappears from central to forward rapidities also shows that the effect is correlated with the nuclear gluon density. In a similar way, we predict that for higher transverse momenta, the away-side peak will reappear, as larger values of $x_A$ will be probed. These modifications of the away-side peak with transverse momenta and rapidity were already predicted in \cite{Marquet:2007vb} as a time where there was no data.

We are not aware of any descriptions of this phenomena that does not invoke saturation effects. We note that apart from our CGC calculation, a successful description based on the KLN saturation model was also recently proposed \cite{Tuchin:2009nf}. While more differential measurements of the coincidence probability, as a function of transverse momentum or rapidity, will provide further tests of our CGC predictions and help understand better this theory of saturation, the analysis of forward di-hadron correlations presented in this work adds further support to the idea that the saturation regime of QCD has been probed at RHIC. Future p+Pb collisions at the LHC will allow definitive tests.


\begin{thebibliography}{99}

\bibitem{RdAu}
  I.~Arsene {\it et al.}  [BRAHMS Collaboration],
  Phys.\ Rev.\ Lett.\  {\bf 93}, 242303 (2004);
  J.~Adams {\it et al.}  [STAR Collaboration],
  Phys.\ Rev.\ Lett.\  {\bf 97}, 152302 (2006).

\bibitem{Albacete:2010bs}
  J.~L.~Albacete and C.~Marquet,
  Phys.\ Lett.\  B {\bf 687}, 174 (2010).

\bibitem{alternatives}
  B.~Z.~Kopeliovich, J.~Nemchik, I.~K.~Potashnikova, M.~B.~Johnson and I.~Schmidt,
  Phys.\ Rev.\  C {\bf 72}, 054606 (2005);
  L.~Frankfurt and M.~Strikman,
  Phys.\ Lett.\  B {\bf 645}, 412 (2007).

\bibitem{rhicprel}
  E.~Braidot for the STAR collaboration,
  arXiv:1005.2378;
  B.~A.~Meredith for the PHENIX Collaboration,
  PoS {\bf DIS2010}, 081 (2010).

\bibitem{Albacete:2010pg}
  J.~L.~Albacete and C.~Marquet,
  Phys.\ Rev.\ Lett.\  {\bf 105}, 162301 (2010).

\bibitem{Dominguez:2011wm}
  F.~Dominguez, C.~Marquet, B.~W.~Xiao and F.~Yuan,
  arXiv:1101.0715 [hep-ph].

\bibitem{othercalc}
  J.~Jalilian-Marian and Y.~V.~Kovchegov,
  Phys.\ Rev.\  D {\bf 70}, 114017 (2004)
  [Erratum-ibid.\  D {\bf 71}, 079901 (2005)];
  N.~N.~Nikolaev, W.~Schafer, B.~G.~Zakharov and V.~R.~Zoller,
  Phys.\ Rev.\  D {\bf 72}, 034033 (2005);
  R.~Baier, A.~Kovner, M.~Nardi and U.~A.~Wiedemann,
  Phys.\ Rev.\  D {\bf 72}, 094013 (2005).

\bibitem{Marquet:2007vb}
  C.~Marquet,
  Nucl.\ Phys.\  A {\bf 796}, 41 (2007).

\bibitem{Dumitru:2010ak}
  A.~Dumitru and J.~Jalilian-Marian,
  Phys.\ Rev.\  D {\bf 82}, 074023 (2010).

\bibitem{loBK}
  I.~Balitsky,
  Nucl.\ Phys.\  B {\bf 463}, 99 (1996);
  Y.~V.~Kovchegov,
  Phys.\ Rev.\  D {\bf 60}, 034008 (1999).

\bibitem{rcBK}
  I.~Balitsky,
  Phys.\ Rev.\  D {\bf 75}, 014001 (2007);
  Y.~V.~Kovchegov and H.~Weigert,
  Nucl.\ Phys.\  A {\bf 784}, 188 (2007);
  J.~L.~Albacete and Y.~V.~Kovchegov,
  Phys.\ Rev.\  D {\bf 75}, 125021 (2007).

\bibitem{Tuchin:2009nf}
  K.~Tuchin,
  Nucl.\ Phys.\  A {\bf 846}, 83 (2010).

\end{thebibliography}
\end{document}